\documentstyle[aps,pra,epsf,epsfig,twocolumn]{revtex} 
\begin{document}

\title{ Fast partial decoherence of \\
        a superconducting flux qubit in a spin bath
}

\author{ Jacek Dziarmaga }

\address{
Intytut Fizyki Uniwersytetu Jagiello\'nskiego, 
ul. Reymonta 4, 30-059 Krak\'ow, Poland 
}

\date{ October 6, 2004 }

\maketitle

\begin{abstract}
The superconducting flux qubit has two quantum states with
opposite magnetic flux. Environment of nuclear spins can find
out the direction of the magnetic flux after a decoherence time
$\tau_0$ inversely proportional to the magnitude of the flux and
the square root of the number of spins. When the Hamiltonian of the
qubit drives fast coherent Rabi oscillations between the states
with opposite flux, then flux direction is flipped at a constant
rate $\omega$ and the decoherence time $\tau=\omega\tau_0^2$ is
much longer than $\tau_0$. However, on closer inspection decoherence
actually takes place on two timescales. The long time $\tau$
is a time of full decoherence but a part of quantum coherence is lost
already after the short time $\tau_0$. This fast partial decoherence
biases coherent flux oscillations towards the initial flux direction
and it can affect performance of the superconducting devices as qubits.
\end{abstract}


Quantum computers can perform certain tasks exponentially faster than
their classical counterparts \cite{QC}. The basic unit of the quantum
computer is a quantum bit or qubit. Several physical implementations
of the qubit were proposed including ion traps \cite{ion}, nuclear magnetic
resonance \cite{NMR}, quantum dots \cite{qdots}, single photon \cite{photon},
and two flavors of superconducting Josephson junction qubits \cite{review}.
The solid state implementations are in principle most easily scallable
and most compatible with the existing classical electronics,
but at the same time they are the most endangered by quantum decoherence.
However, recent experiments \cite{Rabi,Rabi2} in both types of
superconducting qubits show hundreds of coherent oscillations.
This is still far from the minimal requirements of the fault tolerant
quantum computation \cite{FTQC} but it makes superconducting qubits
attractive candidates for scallable quantum technology. In order
to push their performance even further it is essential to understand
their quantum decoherence better.

\section{ Superconducting flux qubit }

The superconducting flux qubit is a superconducting loop broken by
one or more Josephson junctions \cite{review}. The junctions are designed
so that the lowest two quantum eigenstates are states with opposite
magnetic flux. The flux can range from a tiny $10^{-3}\dots 10^{-2}$
fraction of the flux quantum \cite{Delft} to $1/2$ of the quantum
\cite{SB}. The two states span a two-dimensional Hilbert space of the
qubit. I assume convention that the state $|1\rangle~(|0\rangle)$ with
positive(negative) flux is the $+1(-1)$ eigenstate of $\sigma_z$. Degeneracy
of these two states is removed by coherent tunneling between the
opposite flux states driven by the qubit Hamiltonian:
\begin{equation}
H_{Q}~=~\frac12~\omega~\sigma_x~+~\mu~\sigma_z~.
\label{HQ}
\end{equation}
For $\mu=0$ the eigenstates of $H_Q$ are the coherent superpositions of
opposite magnetic fluxes
$
|\pm\rangle=\frac{|1\rangle \pm |0\rangle}{\sqrt{2}}.
$
These Schr\"odinger cats were observed in two independent experiments
\cite{Delft,SB}. For $\mu=0$ the qubit Hamiltonian (\ref{HQ}) drives
coherent oscillations between the states with opposite flux. For example,
the initial state $|1\rangle$ evolves into
$|1\rangle\cos(\frac12\omega t)-i|0\rangle\sin(\frac12\omega t)$ with an
expectation value of the flux operator
\begin{equation}
F(t)~\equiv~\langle~\sigma_z~\rangle~=~\cos(\omega t)
\label{osc}
\end{equation}
oscillating with the Rabi frequency $\omega$. Damped flux oscillations
were observed in a number of experiments \cite{Rabi,Rabi2} in both types
of the superconducting qubit.

The qubit Hamiltonian (\ref{HQ}) is sufficient to perform any one-qubit
operation. Two-qubit operations are possible thanks to inductive coupling
between any two qubits \cite{review}. Entangled eigenstates of two
qubits were detected in Ref.\cite{entangled} and coherent Rabi oscillations
in a system of two inductively coupled qubits were observed in
Ref.\cite{Rabi2}. It is possible to perform single-qubit NOT and two-qubit
CNOT operations in an adiabatic way \cite{adiabatic} that, in particular,
precludes excitation beyond the truncated two-dimensional Hilbert space of
the qubit. However, before it comes to quantum computation quantum
decoherence has to be overcome first.
\section{ Qubit in a spin bath }
Various sources of decoherence for flux qubits have been discussed in
Ref.\cite{estimates}. At relatively high temperatures the main source of
decoherence are normal state quasiparticles. However, density of
quasiparticles is exponentially suppressed at temperatures much less than
the critical temperature. Other decoherence mechanisms including
electromagnetic radiation from the qubit or ohmic dissipation in the
environment are respectively negligible or tractable. Nuclear spins
are argued to play a minor role. Due to relaxation the spins randomly flip
their polarization. Random spins are a source of random magnetic
field which couples to the magnetic moment of the qubit and randomizes
its quantum state. This picture is further corroborated in
Ref.\cite{smirnov} where the spins are assumed to be mutually
non-interacting but each of them is coupled to a bosonic environment.  
The decoherence time is estimated in the range of miliseconds.
The final Eq.(47) in Ref.\cite{smirnov} shows that in the
limit of vanishing nuclear spin relaxation rate ($\gamma_i$ in
Ref.\cite{smirnov}) the decoherence time tends to infinity. As expected,
the external magnetic noise vanishes for vanishing nuclear spin
relaxation rate.

However, apart from generating magnetic noise nuclear spins can be
also silent witnesses of the quantum state of the qubit. In the limit of
vanishing spin relaxation and for negligible spin-spin interaction each
spin is simply precessing in the magnetic field of the qubit. The
direction of the field and the direction of the precession depend on the
state of the qubit. This way the spins can learn the quantum state of the
qubit. Once they know the state any quantum coherence between the states
with opposite flux is lost. This elementary argument shows that decoherence
does not vanish for vanishing spin relaxation.

In this paper I neglect nuclear spin relaxation. When the relaxation rate
is longer than any other relevant timescale, like the frequency of the Rabi
oscillations or frequency of the spin precession in the magnetic field of
a qubit, then this is a very reasonable assumption. Spin lattice relaxation
is also neglected in the paper of Prokofev and Stamp \cite{PS}. In contrast
to the very general formalism employed in Ref.\cite{PS}, in this paper I
apply most elementary methods to a simple model Hamiltonian. This simple
approach benefits with clear interpretation of the results and additional
insights into dynamics of the decoherence process. In particular, the simple
model makes it very clear that even in the absence of any spin lattice
relaxation the spin environment decoheres quantum state of a single qubit.
The decoherence is not an artifact of ensemble average over different
static spin configurations or an ensemble of qubits, as sometimes
claimed in the literature, but a result of genuine entanglement between
the qubit and the spins. 

The qubit interacting with spins is described by a Hamiltonian
\begin{equation}
H~=~H_Q~+~V~+~H_S~.
\nonumber
\end{equation}   
Here $V$ is interaction between the qubit and $N$ spins
\begin{equation}
V~=~\sigma_z~\sum_{n=1}^N \vec{B}(\vec{r}_n)~\vec{\sigma}^{(n)}~. 
\nonumber
\end{equation}
$\vec{B}(\vec{r}_n)$ is a magnetic field of the qubit in the state
$|1\rangle$ at the position $\vec{r}_n$ of the $n$-th spin.
Direction of the magnetic field is reversed in the state $|0\rangle$ as
is accounted for by the operator $\sigma_z$. $\vec{\sigma}^{(n)}$ is a
vector of Pauli matrices 
$\left(\sigma^{(n)}_x,\sigma^{(n)}_y,\sigma^{(n)}_z\right)$
in the Hilbert space of the $n$-th spin. A unitary transformation in the
Hilbert space of each spin brings the interaction Hamiltonian to a more
convenient form
\begin{equation}
V~=~\sigma_z~\sum_{n=1}^N B_n ~ \sigma^{(n)}_x~. 
\label{HQS}
\end{equation}
Here $B_n=|\vec{B}(\vec{r}_n)|>0$ is the strength of the qubit magnetic
field at the location of the $n$-th spin.

\section{ Exactly solvable model }

When the magnetic field from the qubit is stronger than magnetic fields
from other spins, then spin-spin interaction can be neglected, $H_S=0$
and
\begin{equation}
H~=~
\frac12~\omega~\sigma_x~+~
\sigma_z~
\left(
\mu+
\sum_{n=1}^N B_n ~ \sigma^{(n)}_x~.
\right)
\label{H}
\end{equation}
This exactly solvable spin-spin model was also considered in
Ref.\cite{dobrovitski}. The Hamiltonian (\ref{H}) can be easily
diagonalized. The spin part of any eigenstate is
\begin{equation}
|\vec{s}\rangle~=~|s_1\rangle \dots |s_N\rangle~.
\label{s}
\end{equation}
Here $|s_n\rangle$ is an eigenstate of $\sigma^{(n)}_x$ with an eigenvalue
$s_n\in\{+1,-1\}$. In the subspace of $|\vec s\rangle$ the Hamiltonian
(\ref{H}) reduces to an effective qubit Hamiltonian
\begin{equation}
H(\vec s)~=~\frac12~\omega~\sigma_x~+~\sigma_z~b(\vec s)~
\label{Hs}
\end{equation}
with an effective external ``magnetic field''
$b(\vec s)=\mu+\sum_{n=1}^N B_n s_n$. $H(\vec s)$ has eigenvalues
$\pm\Omega(\vec s)\equiv\pm\sqrt{\frac14\omega^2+b^2(\vec s)}$ with
corresponding eigenstates proportional to
$b|+\rangle\pm[\Omega\mp\frac12\omega]|-\rangle$.

\section{ Pure initial state of spins  }

I open discussion of decoherence with an example where the initial
state is
\begin{equation}
|\psi(0)\rangle~=~
\left(\alpha|+\rangle+\beta|-\rangle\right)~|1_1,\dots,1_N\rangle~.
\label{psi0}
\end{equation}
Each spin is initially in the $+1$ eigenstate of its $\sigma^{(n)}_z$.
More general discussion is postponed to Section VIII,
where I consider an ensemble of pure initial states. However, average
over the ensemble will give the same results as the present example so
all the conclusions of this Section are also valid for the ensemble of
spin states.

The Hamiltonian (\ref{H}) evolves the initial state into
\begin{eqnarray}
|\psi(t)\rangle&=&
\frac{1}{2^{\frac{N}{2}}}
\sum_{\vec s}
\left[ {\cal A}(t,\vec s)|+\rangle + {\cal B}(t,\vec s)|-\rangle \right] 
|\vec s\rangle~,
\nonumber\\
{\cal A}(t,\vec s)&=&
e^{-i\Omega t}~b~ 
\frac{\alpha b-\beta(\frac12\omega-\Omega)}{b^2+(\frac12\omega-\Omega)^2}+
\nonumber\\
&&
e^{+i\Omega t}~b~ 
\frac{\alpha b-\beta(\frac12\omega+\Omega)}{b^2+(\frac12\omega+\Omega)^2}~,
\label{A}\\
{\cal B} (t,\vec s)&=&
-e^{-i\Omega t}~ 
(\frac12\omega-\Omega)~ 
\frac{\alpha b-\beta(\frac12\omega-\Omega)}{b^2+(\frac12\omega-\Omega)^2}+
\nonumber\\
&&
-e^{+i\Omega t}~ 
(\Omega+\frac12\omega)~ 
\frac{\alpha b-\beta(\frac12\omega+\Omega)}{b^2+(\frac12\omega+\Omega)^2}~,
\label{B}
\end{eqnarray}
This state is an entangled state of the qubit and the spins: the state of 
the qubit ${\cal A}(t,\vec s)|+\rangle + {\cal B}(t,\vec s)|-\rangle$
depends on the state of spins $|\vec s\rangle$. The state of the qubit can
be described by reduced density matrix $\rho(t)$ obtained by
taking a partial trace over the spins:
\begin{equation}
\rho(t)\equiv
{\rm Tr}_S |\psi(t)\rangle\langle\psi(t)|=
\frac{1}{2^N}
\sum_{\vec s}
\left(
\begin{array}{ccc}
{\cal A}^* {\cal A} &,& {\cal A} {\cal B}^* \\
{\cal A}^* {\cal B} &,& {\cal B} {\cal B}^*
\end{array}
\right).
\label{rhos}
\end{equation}
This matrix form is given in the basis of $|\pm\rangle$ states of the qubit.
$\rho(t)$ is in general a mixed state as can be most conveniently
measured by its ``quadratic entropy''
$S(t)=1-{\rm Tr}\rho^2(t)\in[0,\frac12]$ introduced in Ref.\cite{PazZurek}.
The initial state is pure, $S(0)=0$, but in general $S(t)$ grows as the
interacting qubit and spins get entangled and the state of the qubit is
loosing its initial coherence.

The most right hand side of Eq.(\ref{rhos}) can be formally interpreted
as an average over all possible states of spins, or over all possible
sets $\vec s$ of $N$ independent random variables $s_n$. In the limit
of large $N$ the effective magnetic field
$b(\vec s)=\mu+\sum_{n=1}^N B_n s_n$ becomes a gaussian variable
with a mean of $\mu$ and a variance of $NB^2~\equiv~\sum_{n=1}^N B_n^2$.
In this limit the sum in Eq.(\ref{rhos}) can be approximated by an
integral,
\begin{equation}
\rho(t)=
\int_{-\infty}^{+\infty}
db
\frac{e^{-\frac{(b-\mu)^2}{2NB^2}}}{\sqrt{2\pi NB^2}}
\left(
\begin{array}{ccc}
{\cal A}^* {\cal A} &,& {\cal A} {\cal B}^* \\
{\cal A}^* {\cal B} &,& {\cal B} {\cal B}^*
\end{array}
\right)~.
\label{rhoint}
\end{equation}
This density matrix is formally an average over different values of the
static magnetic noise $b$ generated by different static spin
configurations. The mixed state of the qubit results from an average over
different pure states of the qubit, each of them evolved from the same
initial state but with a different value of $b$ in the Hamiltonian
(\ref{Hs}). However, as the derivation of Eq.(\ref{rhoint}) demonstrates,
this is not an average over different realizations of the experiment with
different static configurations of nuclear spins, but it is an average
over a superposition of different states of spins present in a single
realization of the experiment. This is genuine decoherence due to
entanglement with the spin bath.

In the following sections special cases of this exact solution are worked
out in more detail. I begin with the simplest case when the frequency of
Rabi oscillations $\omega=0$.

\section{ No Rabi oscillations }

The density matrix (\ref{rhoint}) becomes particularly simple in the
absence of flux oscillations when we can set $\Omega(\vec s)=b(\vec s)$
and
\begin{eqnarray}
&&
{\cal A}~=~
\alpha\cos bt -i\beta\sin bt~, \nonumber\\
&&
{\cal B}~=~
\beta\cos bt -i\alpha\cos bt~. \nonumber
\end{eqnarray}
The products ${\cal A}^*{\cal A}$ or ${\cal A}^*{\cal B}$ in the density
matrix (\ref{rhoint}) contain oscillatory terms proportional to
$e^{\pm 2ibt}$. The average over $b$ in Eq.(\ref{rhoint}) dephases these
oscillations to zero, 
\begin{eqnarray}
\int_{-\infty}^{+\infty}db
\frac{e^{-\frac{(b-\mu)^2}{2NB^2}}}{\sqrt{2\pi NB^2}}
e^{\pm 2ibt}~=~
e^{\pm 2i\mu t}
e^{-2NB^2t^2}~,
\label{average0}
\end{eqnarray}
after decoherence time
\begin{equation}
\tau_0~=~\frac{1}{2\sqrt{NB^2}}~.
\label{tau0}
\end{equation}
It is instructive to write the density matrix (\ref{rhoint}) in the basis of 
$\sigma_z$-eigenstates
\begin{eqnarray}
&& \rho(t)~=~\frac12\times \\
&&
\left(
\begin{array}{ccc}
|\alpha+\beta|^2 &,& 
(\alpha+\beta)(\alpha^*-\beta^*)
e^{-\frac{t^2}{2\tau_0^2}+2i\mu t} \\
(\alpha^*+\beta^*)(\alpha-\beta)
e^{-\frac{t^2}{2\tau_0^2}-2i\mu t} &,&
|\alpha-\beta|^2
\end{array}
\right)~.
\label{rhoomega0}
\end{eqnarray}
The off-diagonal coherences between the $\sigma_z$-eigenstates are
destroyed after the decoherence time of $\tau_0$ when the density matrix
becomes diagonal in this basis. This is not quite surprising because
the $\sigma_z$ eigenstates are eigenstates of both the interaction
Hamiltonian $V$ and the qubit Hamiltonian ($\omega=0$) - they are the
ideal pointer states of Ref.\cite{pointers}. Quadratic entropy
of the state (\ref{rhoomega0}) grows like
\begin{equation}
S(t)~=~
\frac12|\alpha+\beta|^2|\alpha-\beta|^2
\left(1-e^{-\frac{t^2}{\tau_0^2}}\right)~.
\label{Somega0}
\end{equation}
The only case when the state of the qubit remains pure, or $S(t)=0$, is when 
the initial state is one of the $\sigma_z$-eigenstates ($\alpha=\pm\beta$).
This is another fundamental property of the pointer states.

\section{ Fast Rabi oscillations }

When Rabi oscillations are much faster than the interaction with the spin
bath, then we have a small parameter
\begin{equation}
\epsilon^2~=~\frac{4NB^2}{\omega^2}~\ll~1~.
\label{epsilon2}
\end{equation}
The matrix elements in the density matrix (\ref{rhoint}) contain oscillatory 
terms which can be approximated as
\begin{eqnarray}
e^{\pm 2i \Omega t}=
e^{\pm 2it \sqrt{\frac14\omega^2+b^2}}\approx
e^{\pm i\omega t
\left[
1+
\frac{4\mu}{\omega_\mu}\frac{(b-\mu)}{\omega_\mu}+
\frac{2\omega^2}{\omega^2_\mu}\frac{(b-\mu)^2}{\omega^2_\mu}
\right]}~
\label{eiOmega}
\end{eqnarray}
with $\omega^2_\mu=\omega^2+4\mu^2$. In this expansion I use the
assumption (\ref{epsilon2}) that
$\frac{(b-\mu)}{\omega_\mu}\simeq\epsilon\ll 1$.

Assuming further that the bias $\mu$ is weak as compared to $\omega$,
\begin{equation}
\mu^2~\ll~\omega^2~,
\end{equation}
the density matrix can be approximated by its leading order term
in both $\mu$ and $(b-\mu)$:
\begin{equation}
\rho(t)~=~
\left(
\begin{array}{ccc}
|\alpha|^2 &,&
\alpha\beta^* z^*(t) e^{-i\omega t}  \\
\alpha^*\beta z(t)   e^{i\omega t} &,&
|\beta|^2
\end{array}
\right)~
\label{rhoRabi}
\end{equation}
with the coherence function
\begin{eqnarray}
z(t)&=&
\frac{\exp\left(-\frac{t^2}{2\tau^2_\mu(1-i\frac{t}{\tau})}\right)}
     {\left(1-i\frac{t}{\tau}\right)^{1/2}}~.
\label{z}
\end{eqnarray}
This function comes from an average of the approximate (\ref{eiOmega})
over the gaussian variable $b$. $z(t)$ has two time scales
\begin{eqnarray}
\tau&=&
\frac{\tau_0}{\epsilon}~
\ll~\tau_0~,
\label{tau}\\
\tau_\mu &=&
\frac{\tau_0}{\left(\frac{2\mu}{\omega}\right)}
~\ll~\tau_0~,
\label{taumu}
\end{eqnarray}
both of them are much longer than the decoherence time $\tau_0$ in the
absence of Rabi oscillations. The coherence function $z(t)$ determines
both the entropy growth
\begin{eqnarray}
S(t)&=&
2 |\alpha|^2 |\beta|^2
\left(1-|z(t)|^2 \right)~,
\label{Sz}
\end{eqnarray}
and coherent flux oscillations
\begin{eqnarray}
F(t)&=&
{\rm Tr} \rho(t)\sigma_z=
\frac12\left[z(t)e^{i\omega t}+{\rm c.c.}\right]~.
\label{Fz}
\end{eqnarray}
starting from the initial $+1$ eigenstate $\sigma_z$. The exact
solution (\ref{rhoRabi}) can be best understood in limiting cases.

For times $t\gg\tau$ the coherence function can be approximated by
$z(t\gg\tau)=\sqrt{\frac{\tau}{t}}
\exp i\left(\frac{\pi}{4}-\frac{\tau}{2\tau^2_\mu}t\right)$ and the
coherence decay is characterized by the power laws:
\begin{eqnarray}
S(t\gg\tau)&=&
2 |\alpha|^2 |\beta|^2
\left(1-\frac{\tau}{t}\right)~,
\label{Stau}\\
F(t\gg\tau)&=&
\sqrt{\frac{\tau}{t}}
\cos\left[\frac{\pi}{4}+
          \left(\omega-\frac{\tau}{2\tau^2_\mu}\right)t\right]~.
\label{Ftau}
\end{eqnarray}
Apart from the shift in the frequency of oscillations, in the regime of
$t\gg\tau$ the coherence decay does not depend on $\tau_\mu$. This power
law decay of flux oscillations was also derived in Ref.\cite{dobrovitski}.
Both the entropy growth and the decay of flux oscillations are due to the
decay of the same coherence function $z(t)$ in the density matrix
(\ref{rhoRabi}). Thus, contrary to Ref.\cite{dobrovitski}, it is not
possible to see oscillations after decay of quantum coherence: flux
oscillations are coherent flux oscillations.

When $t\ll\tau$ the coherence function is
$z(t\ll\tau)=\exp\left(-\frac{t^2}{2\tau^2_\mu}\right)$ and the decay of
coherence is gaussian:
\begin{eqnarray}
S(t\ll\tau)&=&
2 |\alpha|^2 |\beta|^2
\left(1-e^{-\frac{t^2}{\tau^2_\mu}}\right)~,
\label{Smu}\\
F(t\ll\tau)&=&
e^{-\frac{t^2}{2\tau^2_\mu}}
\cos\omega t~.
\label{Fmu}
\end{eqnarray}
When the bias $\mu$ is stronger than the influence of spins,
$NB^2\ll\mu^2\ll\omega^2$, then $\tau_\mu\ll\tau$ and the gaussian
decay of coherence is completed before crossover to the power law
decay after $\tau$. For a weak bias, $\mu^2\ll NB^2\ll\omega^2$, the
coherence decay is a power law. The coherence decays most slowly when the
bias $\mu=0$. The power law (\ref{Ftau}) and gaussian (\ref{Fmu}) decays
are consistent with the decays derived in Ref.\cite{Rabenstein} from a
model with a non-markovian external noise, compare also Ref.\cite{Makhlin}.
Here, these results follow from a microscopic description of quantum 
decoherence in a bath of non-interacting spins.

After decoherence the density matrix (\ref{rhoRabi}) becomes diagonal in
the basis of $\sigma_x$-eigenstates. What is more the entropy remains
zero, $S(t)=0$, only when $\alpha=0$ or $\beta=0$ i.e. when the initial
state of the qubit is an eigenstate of $\sigma_x$. This is not quite
surprising because when $H_Q$ dominates over $V$ and $H_S$, then the
eigenstates of $H_Q$ are expected to be the pointer states \cite{pointer2}.

\section{ Early time partial decoherence }

The density matrix (\ref{rhoRabi}) is the leading order term in the
expansion of the exact density matrix (\ref{rhoint}) in powers of
$\frac{b}{\omega}\simeq\epsilon$. In this Section I include higher
order terms. For the sake of simplicity I consider only the optimal
case of zero bias when $\mu=0$.

The products ${\cal A}^*{\cal B}$ etc. in the exact density matrix
(\ref{rhoint}) are expanded up to second order in powers of
$\frac{b}{\omega}$. The oscillatory factors $e^{\pm i\Omega t}$ are
approximated as in Eq.(\ref{eiOmega}). After these two approximations
the gaussian integral over $b$ in Eq.(\ref{rhoint}) gives the
approximate density matrix accurate up to second order in $\epsilon$
\begin{equation}
\rho(t)~=~\left(
\begin{array}{ccc}
P &,&  C^*  \\
C &,& 1-P
\end{array}
\right)~
\label{rhoepsilon}
\end{equation}
with matrix elements
\begin{eqnarray}
P(t)&=&
|\alpha|^2-
\label{P}\\
&&
\frac12\epsilon^2\left(|\alpha|^2-|\beta|^2\right)
\left[
1-\frac{e^{i\omega t}z^3(t)+{\rm c.c.}}{2}
\right]~,
\nonumber\\
C(t)&=&
\alpha^*\beta z(t)e^{i\omega t}+
\nonumber\\
&&
\frac14
\epsilon^2\left(\alpha^*\beta+{\rm c.c.}\right)
\left[2-e^{i\omega t}z^3(t)\right]- 
\nonumber\\
&&
\frac14
\epsilon^2\left[\alpha^*\beta e^{i\omega t}z^3(t)+{\rm c.c.}\right]~.
\label{C}
\end{eqnarray}
Here $P$ (or $1-P$) is a probability to find the qubit in the state
$|+\rangle$ (or $|-\rangle$) and $C$ is quantum coherence between these
states. The $z(t)$ is the function (\ref{z}) with zero bias $\mu=0$
or $\tau_\mu=\infty$.

For $\mu=0$ the function $z(t)$ has only one timescale $\tau$. The regime
$t\gg\tau$ has been considered in the proceeding Section. In the opposite
regime of early time we can approximate $z(t\ll\tau)=1$ in
Eqs.(\ref{P},\ref{C}) and get the entropy
\begin{eqnarray}
&&
S(t\ll\tau)~=   \label{Searly}\\
&&
\epsilon^2(1-\cos\omega t)
\left\{
|\alpha|^4+|\beta|^4-
\left[
(\alpha^*\beta)^2e^{i\omega t}+{\rm c.c.}
\right]
\right\}
\label{Sepsilon}
\end{eqnarray}
No matter what is the initial state, defined by $\alpha$ and $\beta$, this
entropy does not remain zero but fluctuates with an amplitude proportional
to $\epsilon^2$, see Fig.1. This fluctuating entropy means
partial loss of coherence.

This partial early time decoherence also shows in flux oscillations. The
initial state $\alpha=\beta=\frac{1}{\sqrt{2}}$ leads to flux oscillations
\begin{equation}
F(t)~\equiv~{\rm Tr}~ \sigma_z \rho(t)=
\epsilon^2+(1-\epsilon^2)\cos\omega t~.
\label{fluxearly}
\end{equation}
Unlike the coherent flux oscillation between $+1$ and $-1$ in Eq.(\ref{osc})
this flux oscillates between $+1$ and $-1+2\epsilon^2$, see Fig.2. The 
amplitude
of the coherent oscillation is reduced due to the partial decoherence. The
oscillation is also biased towards the initial positive value of $F$.
Similar effect was also predicted recently in the spin-boson model
\cite{Loss}. Here it follows from the simple and explicitly
solvable spin-spin model. The effect seems to be generic feature
of a non-markovian environment. 

Fast Rabi oscillations make the time of full decoherence $\tau$ much
longer than the decoherence time $\tau_0$ without Rabi oscillations. When
the flux oscillates many times before the spins can learn its orientation,
then it takes the spins much longer time to learn the state of the qubit
than in the static case. On a long timescale covering many flux
oscillations the oscillating magnetic field measured by the spins
effectively averages out to zero. This is the key idea of the quantum
bang-bang control introduced in Ref.\cite{bangbang}. However, this
argument does not apply before the first period of oscillation is
completed. On this short timescale the magnetic field of the qubit is
effectively static and during this short time the spins have a chance to
get a rough idea about its orientation. The spins more or less realize
that the initial flux in Eq.(\ref{fluxearly}) is $+1$ and this fuzzy
knowledge biases the following oscillations in $F(t)$ towards this
positive value. The state of the qubit is partially ``collapsed'' towards
the state with positive flux. This qualitative argument becomes more
substantial when we take as an initial state the equal superpositon of
states with opposite flux $|+\rangle$ (or $\alpha=1,\beta=0$) and compare
the two expressions for entropy, one valid for fast Rabi oscillations
(\ref{Sepsilon}) at an early time long before the first period of
oscillations is completed i.e. for $\omega t\ll 1$, and the other in the
absence of any oscillations (\ref{Somega0}). For such early times both
formulas are the same,
\begin{equation}
S(\omega t\ll 1)~=~\frac{t^2}{2\tau^2_0}~,
\end{equation}
demonstrating that the early time partial decoherence takes place
on the timescale $\tau_0$ as if there were no Rabi oscillations
at all.

Fast Rabi oscillations postpone full decoherence after $\tau$ but they
do not prevent partial decoherence already after $\tau_0$. This effect
may affect scallability of the superconducting flux qubit technology.
The partial loss of coherence $\simeq\epsilon^2$ is small in a single
qubit, but it scales together with the number of qubits $N_{\rm qubits}$
as $N_{\rm qubits}\epsilon^2$ and limits the number of qubits to
\begin{equation}
N_{\rm qubits}~\ll~\frac{1}{\epsilon^2}~.
\end{equation}
Obviously, this upper bound does not apply when quantum error correction
is performed much faster than $\tau_0$.

\section{ Ensemble of spins states }
 
So far the initial state of spins was a special pure state. In this
Section I assume the unbiased initial ensemble of all possible spins' 
states described the unit density matrix of spins $1_S$. The initial state 
of the qubit and the spins is
\begin{equation}
\rho_{QS}(0)~=~
\left(\alpha|+\rangle+\beta|-\rangle\right)
\left(\alpha^*\langle+|+\beta^*\langle-|\right)
~\otimes~1_S~.
\nonumber
\end{equation}
The unit matrix $1_S$ can be expressed in any basis of spins states but a 
representation
$1_S=\frac{1}{2^N}\sum_{\vec s}|\vec s\rangle\langle\vec s|$ most
directly leads to the reduced density matrix of the qubit
\begin{equation}
\rho(t)\equiv
{\rm Tr}_S \rho_{QS}(t)=
 \frac{1}{2^N}
\sum_{\vec s}
\left(
\begin{array}{cc}
{\cal A}^* {\cal A} & {\cal A} {\cal B}^* \\
{\cal A}^* {\cal B} & {\cal B} {\cal B}^*
\end{array}
\right).
\nonumber
\end{equation}
This density matrix is identical with the density matrix (\ref{rhos})
obtained from the initial pure state in Eq.(\ref{psi0}). All the results
derived from the pure state pass without any modification to the ensemble.
The initial state $1_S$ can be also interpreted as a high temperature
thermal state of nuclear spins.

\section{ Estimates  }

In this Section I attempt to make some numerical estimates. For example,
in the qubit design considered in Ref.\cite{smirnov} the parameters are
estimated as: $N=10^8$, $B=10^{4\dots 5}{\rm s}^{-1}$ and
$\omega=10^9{\rm s}^{-1}$. These numbers lead to $\tau_0=10^{-8\dots-9}$s,
$\epsilon^2=10^{-2\dots 0}$, and $\tau=10^{-7\dots -9}$s. In the best case
of $\epsilon^2=10^{-2}$ one can see $\omega\tau=\epsilon^{-2}=10^2$ of
coherent flux oscillations before the decoherence time $\tau$. Thus it is
possible to see hundreds of oscillations as in the experiments \cite{Rabi}.
For the same parameters the fast partial decoherence limits the number
of qubits to $N_{\rm qubits}\ll 10^2$. These numbers leave, however, an
encouraging room for current experiments.

\section{ Conclusion }

The environment of spins is a witness to the quantum state of the
superconducting flux qubit. The qubit gets entangled with spins after a
decoherence time $\tau_0$. Coherent flux oscillations driven by the qubit
Hamiltonian postpone full decoherence after a much longer time $\tau$, but
they do not prevent partial decoherence already after $\tau_0$. This
partial decoherence biases the flux oscillations towards the initial
orientation of the flux.

{\bf Acknowledgements.---} I would like to thank Leo Stodolsky and Jacek 
Wosiek for encouragement. This work was supported in part by the the KBN 
grant PBZ-MIN-008/P03/2003. 

\begin{figure}[ht]
\vspace*{2cm}
\epsfig{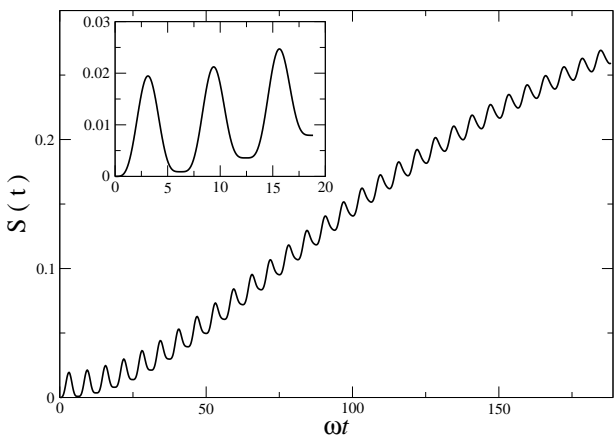}
\vspace*{0.7cm}
\caption{ Entropy $S(t)$ according to the exact Eq.(13)
for $\omega=1$, $\mu=0$, $\epsilon=0.1$, and
$\alpha=\beta=\frac{1}{\sqrt{2}}$. The timescale for full decoherence
is $\tau=400$ and the early time partial decoherence sets in
on the timescale $\tau_0=10$. The inset shows the entropy at early times.
The non-vanishing early time entropy means partial decoherence.
}
\label{Sfigearly}
\end{figure}               

\begin{figure}[ht]
\epsfig{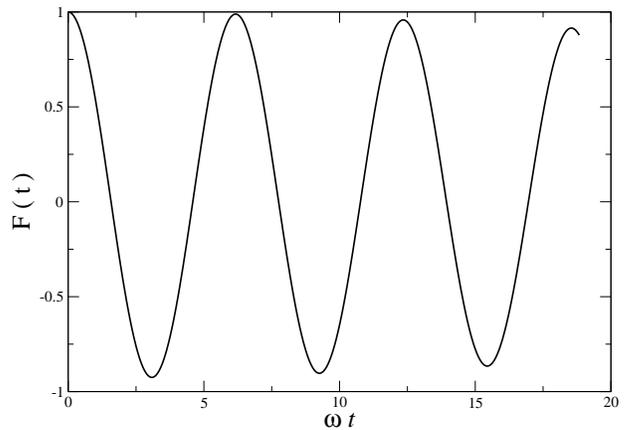}
\vspace*{0.1cm}
\caption{ Flux $F(t)$ according to the exact Eq.(13) for $\omega=1$,
$\mu=0$, $\epsilon=0.2$, and $\alpha=\beta=\frac{1}{\sqrt{2}}$. As a
result of the fast partial decoherence the flux oscillation is biased
towards the initial $F(0)=+1$.
}
\label{Fearly}
\end{figure}


\begin{thebibliography}{99}

\bibitem{QC} M.A.~Nielsen and I.L.~Chuang, 
             {\it Quantum Computation and Quantum Information}, 
             Cambridge University Press, 2001.

\bibitem{ion} D.~Kielpinski, B.E.~King, C.J.~Myatt, et al.,
              Phys.Rev.A {\bf 61}, 032310 (2000). 

\bibitem{NMR} E.~Knill, R.~Laflamme, R.~Martinez, and C.H.~Tseng,
              Nature {\bf 404}, 368 (2000);
              L.M.K.~Vandersypen, M.~Steffen, G.~Breyta, C.S.~Yannoni, 
              M.H.~Sherwood, and I.L.~Chuang,
              Nature {\bf 414}, 883 (2001).

\bibitem{qdots}  R.~Hanson {\it et al.}, cond-mat/0303139;
                 D.~Loss and D.P.~DiVincenzo,
                 Phys.Rev.A {\bf 57}, 120 (1998).

\bibitem{photon} E.~Knill, R.~Laflamme, and G.J.~Milburn,
                 Nature {\bf 409}, 46 (2001). 

\bibitem{review} Y.~Makhlin, G.~Sch\"on and A.~Shnirman, 
                 Rev.Mod.Phys. {\bf 73}, 357 (2001). 

\bibitem{Rabi} Y. Nakamura, Yu.A. Pashkin, and J.S. Tsai,
               Nature {\bf 398}, 786 (1999);
               J.R. Friedman, V. Patel, W. Chen, S.K. Tolpygo, and
               J.E. Lukens, Nature {\bf 406}, 43 (2000);
               C.H. van der Wal, A.C.J. ter Haar, F.K. Wilhelm,
               R.N. Schouten, C. Harmans, T.P. Orlando, S. Lloyd, 
               and J.E. Mooij,
               Science {\bf 290}, 773 (2000); 
               I. Chiorescu, Y. Nakamura, C.J.P.M. Harmans, and 
               J.E. Mooij, 
               Science {\bf 299}, 1869 (2003);
               D. Vion, A. Aassime, A. Cottet, P. Joyez, H. Pothier,
               C. Urbina, D. Esteve, and M.H. Devoret, 
               Science {\bf 296}, 886 (2002);
               J.M. Martinis, S. Nam, J. Aumentado, and C. Urbina,
               Phys.\ Rev.\ Lett. {\bf 89}, 117901 (2002);
               E. Il'ichev, N. Oukhanski, A. Izmalkov, Th. Wagner,
               M. Grajcar, H.-G. Meyer, A. Yu. Smirnov, A.M. van den 
               Brink, M.H.S. Amin, and A.M. Zagoskin, 
               Phys.\ Rev.\ Lett. {\bf 91}, 097906 (2003);
               T. Duty, D. Gunnarsson, K. Bladh, R.J. Schoelkopf, 
               P. Delsing, cond-mat/0305433.

\bibitem{Rabi2} Yu. A. Pashkin, T. Yamamoto, O. Astafiev, Y. Nakamura,
                D.V. Averin, and J.S. Tsai, 
                Nature {\bf 421}, 823 (2003).  

\bibitem{FTQC}  P.W.~Shor, Phys.Rev.A {\bf 52}, 2493 (1995).

\bibitem{Delft} C.H.~van der Wal, A.C.J.~ter Haar, F.K.~Wilhelm,
                R.N.~Schouten, C.J.P.M.~Harmans, T.P.~Orlando, S.~Lloyd,
                and J.E.~Mooij, Science {\bf 290}, 773 (2000).

\bibitem{SB} J.R.~Friedman, V.~Patel, W.~Chen, S.K.~Tolpygo and 
             J.E.~Lukens, Nature {\bf 406}, 43 (2000).

\bibitem{entangled} A.~Izmalkov, M.~Grajcar, E.~Il'ichev, Th.~Wagner, 
                    H.-G.~Meyer, A.Yu.~Smirnov, M.H.S.~Amin, 
                    A.~Maassen van den Brink, A.M.~Zagoskin, 
                    cond-mat/0312332.

\bibitem{adiabatic} V.~Corato, P.~Silvestrini, L.~Stodolsky, J.~Wosiek,
                    Phys.Lett.A {\bf 309}, 206 (2003);
                    Phys.Rev.B {\bf 68}, 224508 (2003).

\bibitem{estimates} T.P.~Orlando, J.E.~Mooij, L.~Tian, 
                    C.H.~van der Wal, L.S.~Levitov, 
                    S.~Lloyd and J.J.~Mazo, 
                    Phys.Rev.B {\bf 60}, 15398 (1999). 

\bibitem{PS} N.~Prokofev and P.~Stamp, cond-mat/0006054;
             Rep.Prog.Phys.{\bf 63}, 669 (2000).

\bibitem{smirnov} G.~Rose and A.Yu.~Smirnov, 
                  J.Phys.: Condens. Matter {\bf 13}, 11027 (2001). 
                       
\bibitem{dobrovitski} V.V.~Dobrovitski, H.A.~De Raedt, M.I.~Katsnelson, 
                      and B.N.~Harmon, 
                      Phys.Rev.Lett. {\bf 90}, 210401 (2003).  

\bibitem{PazZurek} W.H.~Zurek, S.~Habib, and J.P.~Paz,
                   Phys.Rev.Lett. {\bf 70}, 1187 (1993).  

\bibitem{pointers} W.H.~Zurek, Phys.Rev.D {\bf 24}, 1516 (1981);
                               Phys.Rev.D {\bf 26}, 1862 (1982).    

\bibitem{Rabenstein} K.~Rabenstein, V.A.~Sverdlov, and D.V.~Averin,
                     cond-mat/0401519.

\bibitem{pointer2} J.P.~Paz and W.H.~Zurek,
                   Phys.Rev.Lett. {\bf 82}, 5181 (1999).  

\bibitem{Makhlin} Y.~Makhlin and A.~Shnirman,
                  cond-mat/0308297;
                  cond-mat/0312585.

\bibitem{Loss}    D.~Loss and D.P.~DiVincenzo,
                  cond-mat/0304118.

\bibitem{bangbang} L.~Viola and S.~Lloyd, Phys.Rev.A {\bf 58}, 2733 
(1998).

\end{thebibliography}
\end{document}